\documentclass[12pt,fleqn]{article}
\usepackage{amsmath,amssymb}
\usepackage{epsfig}
\usepackage[top=0.75in, bottom=0.75in, left=0.75in, right=0.75in, dvips]{geometry}
\usepackage{subfigure}
\begin{document}

AN INVESTIGATION OF UNIFORM EXPANSIONS OF LARGE ORDER BESSEL
FUNCTIONS IN GRAVITATIONAL WAVE SIGNALS FROM PULSARS\\

F.A. Chishtie$^{\dag}$, K.M. Rao$^{\dag}$, I.S.
Kotsireas$^{\ast}$, S.R. Valluri$^{\ddag}$

Departments of $^{\dag}$Applied Mathematics, $^{\ddag}$Physics \&
Astronomy, University of Western Ontario, London, Canada

$^{\ast}$Department of Physics and Computer Science, Wilfrid
Laurier University, Waterloo, Canada

E-mail: farrukh4ad@yahoo.com, krao@uwo.ca, ikotsire@wlu.ca,
valluri@uwo.ca

\begin{abstract}

In this work, we extend the analytic treatment of Bessel functions
of large order and/or argument. We examine uniform asymptotic
Bessel function expansions and show their accuracy and range of
validity. Such situations arise in a variety of applications, in
particular the Fourier transform of the gravitational wave signal
from a pulsar. The uniform expansion we consider here is found to
be valid in the entire range of the argument.
\end{abstract}

\section{Introduction}
The detection of gravitational waves (GW) from astrophysical
sources is one of the most important problems in experimental
gravitation today. Large laser interferometric gravitational wave
detectors like LIGO, VIRGO, LISA, TAMA 300, GEO 600 and AIGO may
in the near future open a new window for the study of a great
variety of nonlinear curvature phenomena.

In recent works \cite{JVD96,CQG2002,VCV2005} we have analyzed the
Fourier transform (FT) of the Doppler shifted GW signal from a
pulsar with the use of the plane wave expansion in spherical
harmonics (PWESH), which has a variety of applications in many
areas \cite{MWIEEE,KThorne80,AO97}. The consequent analysis of the
Fourier transform of the GW signal from a pulsar has a very
interesting and convenient development in terms of the resulting
spherical Bessel, generalized hypergeometric, gamma and Legendre
functions. Both rotational and orbital motions of the Earth and
spindown of the pulsar can be considered in this analysis which
happens to have a nice analytic representation for the GW signal
in terms of the above special functions. The signal can then be
studied as a function of a variety of different parameters
associated with both the GW pulsar signal as well as the orbital
and rotational parameters. The numerical analysis of this
analytical expression for the signal offers a challenge for fast
and high performance parallel computation. Recent studies of the
Cosmic Microwave Background Explorer have raised the interesting
question of the study of very large multipole moments with angular
momentum $l$ and its projection $m$ going up to very large values
of $l\sim1000$. Such problems warrant an intensive analytic study
supplemented by numerical and parallel computation.

Since our FT depends on the Bessel function, a computational issue
arises due to large values of the index or order $n$ of the
function. In the GW form of the pulsar, the Doppler shifted
orbiting motion gives rise to Bessel functions $J_{n}(\frac{2 \pi
f_0 A \sin \theta}{c})$, where $\frac{2 \pi f_0 A \sin\theta}{c}$
is large for non-negligible angle $\theta$. Even for
$\sin{\theta}\sim\frac{1}{1000}$, the argument is large
necessitating the consideration of large values of $n$. The
motivation of this work is to extend the analysis in Watson
\cite{Watson} for large index, argument and overlapping
situations. Meissel \cite{M1} has made derivations for large order
Bessel functions both when the argument is smaller than the order
and vice versa. These were studied in the GW context in an earlier
work by us \cite{CVRSW}. We extend this study by considering
another type of expansion, also referred to as the ``uniform"
Bessel expansion, which, being an expansion in the Airy functions,
allows for single series form, hence the name \cite{ABS}. As an
application, we will address the phenomenological situation of GW
signal analysis of large order $n$ (which does arise with
combinations of $l$ and $m$).

The expansion in terms of the Bessel functions has recently found
a valuable application \cite{PRIX05} in the identification of
global parameter space correlations of a coherent matched
filtering search for continuous GW from isolated neutron stars.
The authors have done an interesting pioneering analysis of the
global correlations and indicate that the locations of local
maxima of the detection statistics involve Bessel functions and
are dominated by the Doppler shift due to the Earth's orbital
motion. We have made a preliminary extension of their analysis
that makes use of the summation of infinite series involving
Bessel functions and other relevant functions of the parameters
and physical variables. Their various levels of approximation can
be further improved for a comparison with exact numerical results.

Captures of stellar-mass compact objects (CO) by massive black
holes are important capture sources for the Laser Interferometer
Space Antenna (LISA), the space based GW detector due to be
launched in about a decade \cite{PM}. Higher Harmonics of the
orbital frequency of the COs arise in the post Newtonian (PN)
capture GW model forms and contribute considerably to the total
signal to noise (S/N) ratio of the waveform. The GW form can be
decomposed into gravitational multipole moments which are treated
in the Fourier analysis of Keplerian eccentric orbits. The
radiation depends strongly on the orbital eccentricity $e$, and
Bessel functions $J_{n}(ne)$ are a natural consequence of the
analysis. Such calculations involving gravitational wave forms
necessitate a very accurate evaluation of Bessel functions of
large order and argument due to the importance of orbital
eccentricity. Accurate and efficient evaluations of Bessel
functions for constructing reliable gravitational wave templates
from binary stars with arbitrary orbital eccentricity have been
done by Pierro et al. \cite{Pierro} using generalized
Carlini-Meissel expansions. The methods investigated in the
current work can also be applied to such calculations relating to
LISA capture sources.

The calculation of partial derivatives of the potential scattering
phase shifts which often contain Bessel and Legendre functions of
large order angular momentum $l$, with respect to angular momentum
arise in a variety of scattering problems in atomic, molecular and
nuclear physics. In particular, large values of $l$ can arise in
rainbow, glory and orbit scattering \cite{BM72,FW59}, in black
hole gravitational fields \cite{RW57}, and in catastrophe theory
\cite{PS78}. The analysis in our paper should help provide
suitable approximations for large order and/or argument for the
Bessel functions that arise in such problems.

In section 2 we give the relevant expressions from our work on GW
signal analysis to demonstrate the occurrence of large order
Bessel functions. Section 3 presents the uniform expansion and the
expressions associated with it. We discuss the results in Section
4 and compare them with other expansions and the exact values of
the Bessel functions.

\section{Fourier Transform of the GW signal}
The FT for the GW Doppler shifted pulsar signal \cite{JVD96} is
given as follows:

\begin{eqnarray}
\widetilde{h}(f)=S_{n l
m}(\theta,\phi)={\sum_{n=-\infty}^{\infty}\sum_{l=0}^{\infty}\sum_{m=-l}^{l}\psi_0
\psi_1 \psi_2 \psi_3 \psi_4}
\end{eqnarray}
where
\begin{equation}
\psi_0(n,l,m,\theta,\phi)= 4\pi i^{l}Y_{l m}(\theta ,\phi )N_{l
m}P_{l }^{m}(\cos \alpha )
\end{equation}
\begin{eqnarray}
\psi_1(n,\theta, \phi)=T_{rE}\sqrt{\frac{\pi}{2}}e^{-i\frac{2\pi
f_{0}A}{c}\sin \theta \cos \phi } i^{n}e^{-in\phi}
J_{n}\left(\frac{2\pi f_{0}A\sin \theta }{c}\right)
\end{eqnarray}
\begin{equation}
\psi_2(l,n,m)=\left\{\frac{1-e^{i\pi (l -B_{orb})R}}{1-e^{i\pi (l
-B_{orb})}} \right\} \frac{e^{-iB_{orb}\frac{\pi}{2}}}{2^{2l}}
\end{equation}

\begin{eqnarray}
\psi_3(l,n,m)=k^{l+\frac{1}{2}}\frac{\Gamma \left(l +1\right)
}{\Gamma \left(l +\frac{3}{2}\right)\Gamma \left(\frac{l
+B_{orb}+2}{2}\right)\Gamma \left(\frac{l -B_{orb}+2}{2}\right)}
\end{eqnarray}

\begin{eqnarray}
\psi_4(l,n,m)=_1F_{3}\left(l+1;l+\frac{3}{2};\frac{l
+B_{orb}+2}{2},\frac{l -B_{orb}+2}{2};\frac{-k^{2}}{16}\right)
\end{eqnarray}

Here $\alpha$ is the co-latitude detector angle and angles
$\theta$ and $\phi$ specify the direction of the pulsar source.
Also $\omega_0=2\pi f_0$, $\omega_{orb}=\frac{2\pi}{T_{orb}}$
($T_{orb}=365$ days), $\omega_{rot}=\frac{2\pi}{T_{rE}}$
($T_{rE}=1$ day),
$B_{orb}=2\left(\frac{\omega-\omega_0}{\omega_{rot}}+\frac{m}{2}+\frac{n
\omega_{orb}}{\omega_{rot}}\right)$, $k=\frac{4\pi f_0 R_E
\sin(\alpha)}{c}$ ($R_E$ is the radius of Earth, $c$ is the
velocity of light) and $A=1.5 \times 10^{11}$ meters is the
Sun-Earth distance.

\section{Uniform expansion of the Bessel function}

An expansion that approximates the Bessel function $J_{\nu}(\nu
z)$ for large $\nu$, which in principle is valid for all real
values of $z$, is given in \cite{ABS} as

\begin{equation}
J_{\nu}(\nu
z)\sim\left(\frac{4\zeta}{1-z^2}\right)^{1/4}\left[\frac{Ai(\nu^{2/3}\zeta)}{\nu^{1/3}}\sum_{k=0}^{\infty}
\frac{a_k(\zeta)}{\nu^{2k}}+\frac{Ai'(\nu^{2/3}\zeta)}{\nu^{5/3}}\sum_{k=0}^{\infty}
\frac{b_k(\zeta)}{\nu^{2k}}\right]
\label{uniform}
\end{equation}

\noindent where
\begin{equation}
a_k(\zeta)=\sum_{s=0}^{2k}\mu_s \zeta^{-3s/2} u_{2k-s}\left(1-z^2
\right)^{-1/2}
\end{equation}

\begin{equation}
b_k(\zeta)=-\zeta^{-1/2}\sum_{s=0}^{2k+1}\lambda_s \zeta^{-3s/2}
u_{2k-s+1}\left(1-z^2 \right)^{-1/2}
\end{equation}

\begin{equation}
\lambda_s=\frac{(2s+1)(2s+3)\ldots(6s-1)}{s!(144)^s}
\end{equation}

\begin{equation}
\mu_s=\frac{6s+1}{1-6s}\lambda_s
\end{equation}

\begin{equation}
\frac{2}{3}\zeta^{3/2}=\ln\left(\frac{1+\sqrt{1-z^2}}{z}\right)-\sqrt{1-z^2}
\label{xlessv}
\end{equation}

\begin{equation}
\frac{2}{3}(-\zeta)^{3/2}=\sqrt{z^2-1}-\arccos\left(\frac{1}{z}\right)
\label{xgreaterv}
\end{equation}

$Ai(x)$ represents the Airy function. Equations \ref{xlessv} and
\ref{xgreaterv} are used to define $\zeta$, the choice depending
on the region of interest. Equation \ref{xlessv} is used when the
argument of the Bessel function is less than the order, while
equation \ref{xgreaterv} is used when the argument is larger than
the order. Also, the $u_i$ are defined recursively
by \\

$u_{0}(t) = 1$

$u_{k+1}(t) = \frac{1}{2} t^2 \left( 1 - t^2 \right) u_{k}'(t) +
\frac{1}{8} \int_0^t \left( 1 - 5t^2 \right) u_{k}(t) dt$
\hspace{25pt} ($k = 0, 1, 2, ...$)\\

Using the symbolic package MAPLE, we have computed these functions
up to $u_{11}$, and the results are given below. They may,
however, be calculated in principle up to any order, and MAPLE can
perform this task fairly rapidly.

\begin{eqnarray}
u_1&=-\frac {5}{24}{t}^{3}+\frac{1}{8}t \nonumber
\end{eqnarray}
\begin{eqnarray}
u_2&=\frac{385}{1152}{t}^{6}-\frac{77}{192}{t}^{4}+\frac{9}{128}{t}^{2}\nonumber
\end{eqnarray}
\begin{eqnarray}
u_3&=-\frac {85085}{82944}{t}^{9}+\frac
{17017}{9216}{t}^{7}-\frac{4563}{5120}{t}^{5}+\frac
{75}{1024}{t}^{3}\nonumber
\end{eqnarray}
\begin{eqnarray}
u_4&=\frac {37182145}{7962624}{t}^{12}-\frac
{7436429}{663552}{t}^{10}+\frac{144001}{16384}{t}^{8}
-\frac{96833}{40960}{t}^{6}+\frac{3675}{32768}{t}^{4}\nonumber
\end{eqnarray}
\begin{eqnarray}
u_5&=-\frac {5391411025}{191102976}{t}^{15}+\frac {5391411025}{
63700992}{t}^{13}-\frac {108313205}{1179648}{t}^{11} +\frac
{250881631}{5898240}{t}^{9}-\frac {67608983}{9175040}{t}^{7}+
\frac {59535}{262144}{t}^{5}\nonumber
\end{eqnarray}
\begin{eqnarray}
u_6&=\frac{5849680962125}{27518828544}{t}^{18}-\frac{1169936192425}{1528823808}{t}^{16}+\frac{4445922195}{4194304}{t}^{14}
-\frac{33010308331}{47185920}{t}^{12}+\frac
{1441372804469}{6606028800}{t}^{10}-\nonumber
\end{eqnarray}
\begin{eqnarray}
&\frac {388895895}{14680064}{t}^{8}+\frac {2401245}{
4194304}{t}^{6}\nonumber
\end{eqnarray}
\begin{eqnarray}
u_7&=-\frac {1267709431363375}{660451885056}{t}^{21}+\frac {
1774793203908725}{220150628352}{t}^{19}-\frac {36927006432745}{
2717908992}{t}^{17}+\frac {10559432785187}{905969664}{t}^{15}-
\nonumber
\end{eqnarray}
\begin{eqnarray}
&\frac {1602251736839}{301989888}{t}^{13}+\frac{1007390378503}{
838860800}{t}^{11}-\frac {25388505925}{234881024}{t}^{9}+\frac
{57972915}{33554432}{t}^{7}\nonumber
\end{eqnarray}
\begin{eqnarray}
u_8&=\frac
{2562040760785380875}{126806761930752}{t}^{24}-\frac{512408152157076175}{5283615080448}{t}^{22}+\frac{75358832548684685}{391378894848}{t}^{20}-\frac{39803268297948155}
{195689447424}{t}^{18}+ \nonumber
\end{eqnarray}
\begin{eqnarray}
&\frac {3542717254441859}{28991029248}{t}^{16}-\frac
{276439228010667}{6710886400}{t}^{14}+\frac{667955999804539}{93952409600}{t}^{12}-\frac
{928090660435}{1879048192}{t}^{10}+\frac
{13043905875}{2147483648}{t}^{8} \nonumber
\end{eqnarray}
\begin{eqnarray}
u_9&=-\frac{6653619855759634132375}{27390260577042432}{t}^{27}+\frac
{1330723971151926826475}{1014454095446016}{t}^{25}-\frac {
3128960418491082175}{1043677052928}{t}^{23}+\nonumber
\end{eqnarray}
\begin{eqnarray}
&\frac { 35348759075759093965}{9393093476352}{t}^{21}-\frac {
17618708259302571707}{6262062317568}{t}^{19}+\frac {
817138105244771959}{644245094400}{t}^{17}-\frac {
3739063570455884033}{11274289152000}{t}^{15}+\nonumber
\end{eqnarray}
\begin{eqnarray}
&\frac {1359491937582325}{30064771072}{t}^{13}-\frac
{472414367256615}{188978561024}{t}^{11}+\frac
{418854310875}{17179869184}{t}^{9}\nonumber
\end{eqnarray}
\begin{eqnarray}
u_{10}&=\frac{4318199286388002551911375}{1314732507698036736}{t}^{30}-\frac
{4318199286388002551911375}{219122084616339456}{t}^{28}+
\frac{91891691150784941691275}{1803473947459584}{t}^{26}-\nonumber
\end{eqnarray}
\begin{eqnarray}
&\frac {16705838021516291703055}{225434243432448}{t}^{24}+\frac {
19941766574064397067317}{300578991243264}{t}^{22}-\frac {
5227733363217471800551}{139156940390400}{t}^{20}+\nonumber
\end{eqnarray}
\begin{eqnarray}
&\frac { 1232816120293110459821}{92771293593600}{t}^{18}-\frac {
35892416277828185849}{12884901888000}{t}^{16}+\frac {
45660648644355162105}{148159191842816}{t}^{14}-\frac {
20993386079260455}{1511828488192}{t}^{12}\nonumber
\end{eqnarray}
\begin{eqnarray}
&+\frac {30241281245175}{ 274877906944}{t}^{10} \nonumber
\end{eqnarray}
\begin{eqnarray}
u_{11}&=-\frac{1556514560957130010757145625}{31553580184752881664}{t}^{33}+\frac
{3424332034105686023665720375}{10517860061584293888}{t}^{31}-\frac{365967912305800454531456575}{389550372651270144}{t}^{29}\nonumber
\end{eqnarray}
\begin{eqnarray}
&+\frac{201734750392525792544487385}{129850124217090048}{t}^{27}-\frac{11694306169843138084657687}{7213895789838336}{t}^{25}+
\frac{164293183874328160710877}{148434069749760}{t}^{23}\nonumber
\end{eqnarray}
\begin{eqnarray}
&-\frac{23186185730591085896097833}{46756731971174400}{t}^{21}+\frac{527174389121818780771231}{3710851743744000}{t}^{19}-\frac{329641577686894230674187}{13469017440256000}{t}^{17}+\nonumber
\end{eqnarray}
\begin{eqnarray}
&\frac{241770821762631191867}{107752139522048}{t}^{15}-\frac{26416375998266454375}{314460325543936}{t}^{13}+\frac{1212400457192925}{2199023255552}{t}^{11}\nonumber
\end{eqnarray}
\\
In Table \ref{table: nu300} we compare the values generated by
this expansion with the exact Bessel function values, as well as
the other expansions (Watson's expansion, epsilon expansion and
Meissel's third expansion) from our previous work \cite{CVRSW} for
order $\nu = 300$. (Refer to the discussion section for the
formulae for these expansions.) We also give a comparison for a
larger order, $\nu = 50000$, in Table \ref{table: nu50000}. We
were not able to give exact values of the Bessel function for $\nu
= 50000$ due to the immense computational power required, but we
give values for the other expansions for comparison.

\clearpage

\begin{table}%[ht]
\caption{Comparison of expansions for $J_{\nu}(x)$ with $\nu =
300$.} \vspace{10pt} \centering \small
\begin{tabular}{|c|c|c|c|c|c|c|} \hline \textbf{\textit{x}} & \textbf{290} &
\textbf{295} & \textbf{298} & \textbf{302} & \textbf{305} &
\textbf{310}\\ \hline \textbf{Watson} & 0.007669 & 0.027187 &
0.048992 & 0.084432 &  0.100255 & 0.057368\\
\hline \textbf{Epsilon} & 0.0076775 & 0.027215 & 0.049028 &
0.084335 & 0.10014& 0.057419
\\\hline \textbf{Uniform} & 0.007677 &
0.027213 & 0.27573 &  -0.083138  &  0.10014 &  0.057419\\\hline
\textbf{Exact} & 0.007677 & 0.027215 & 0.049028 & 0.084335 &
0.10014  & 0.057419\\\hline
\end{tabular}\\
\label{table: nu300}
\end{table}

\begin{table}%[ht]
\caption{Comparison of expansions for $J_{\nu}(x)$ with $\nu =
50000$.} \vspace{10pt} \centering \small
\begin{tabular}{|c|c|c|c|c|c|c|} \hline \textbf{\textit{x}} & \textbf{49800} &
\textbf{49900} & \textbf{49999} & \textbf{50001} &
\textbf{50100} & \textbf{50200}\\
\hline \textbf{Watson} & $3.84377\times 10^{-8}$ & $1.02742\times
10^{-4}$ &
0.011802 & 0.012444 & -0.013648 & 0.001398\\
\hline \textbf{Epsilon} &  80.294 & 0.0016699 & 0.011839 &  0.012444 & -0.012451 & -6.3189\\
\hline \textbf{Uniform} & $3.8445\times 10^{-8}$ & 0.00010275 &
$8.7321\times 10^{13}$ & $3.8936\times 10^{11}$ & -0.013648 &
0.001398 \\ \hline \textbf{Exact} & N/A & N/A & N/A & N/A & N/A &
N/A \\\hline
\end{tabular}\\
\label{table: nu50000}
\end{table}

The uniform expansion values were calculated using four terms in
the series in equation \ref{uniform}. We note that there is a
significant discrepancy between the uniform expansion values and
the exact values close to the transition region. Temme
\cite{Temme} has noted that this is due to numerical singularities
rather than to analytic singularities, and he has given expansions
of the functions $a_k(\zeta)$ and $b_k(\zeta)$ appearing in
equation \ref{uniform}. Using these improved formulae, we obtain
the following uniform expansion values.

\begin{table*}[h]
\centering \small
\begin{tabular}{|c|c|c|c|c|c|c|} \hline \textbf{\textit{x}} & \textbf{290} &
\textbf{295} & \textbf{298} & \textbf{302} & \textbf{305} &
\textbf{310}\\ \hline \textbf{$\nu=300$} & 0.00767703 & 0.027215 &
0.049028 & 0.084335 & 0.100143 & 0.057419 \\
\hline \textbf{\textit{x}} & \textbf{49800} & \textbf{49900} &
\textbf{49999} & \textbf{50001} & \textbf{50100} &
\textbf{50200}\\ \hline \textbf{$\nu=50000$} & $3.844421\times
10^{-8}$
& 0.00010275 & 0.011839 & 0.012444 & -0.013648 & 0.0013977 \\
\hline
\end{tabular}
\end{table*}

A similar expansion for the first derivative of the Bessel
function is also given in \cite{ABS}. This expression is given
below.

\begin{equation}
J_{\nu}^{\prime} (\nu z) \sim -\frac{2}{z}
\left(\frac{1-z^2}{4\zeta} \right)^{\frac{1}{4}} \left[
\frac{Ai(\nu^{2/3} \zeta)}{\nu^{4/3}} \sum_{k=0}^{\infty}
\frac{c_k (\zeta)}{\nu^{2k}} + \frac{Ai^{\prime} (\nu^{2/3}
\zeta)}{\nu^{2/3}} \sum_{k=0}^{\infty} \frac{d_k
(\zeta)}{\nu^{2k}} \right]
\end{equation}

\noindent where
\begin{equation}
c_k (\zeta) = - \zeta^{\frac{1}{2}} \sum_{s=0}^{2k+1} \mu_{s}
\zeta^{-3s/2} v_{2k-s+1}\left[ (1-z^2)^{-\frac{1}{2}} \right]
\end{equation}

\begin{equation}
d_k (\zeta) = \sum_{s=0}^{2k} \lambda_{s} \zeta^{-3s/2}
v_{2k-s}\left[ (1-z^2)^{-\frac{1}{2}} \right] .
\end{equation}

The coefficients $v_i$, like $u_i$, are defined recursively,
according to the following relation.
\\

$v_{0}(t) = 1$

$v_{k}(t) = u_{k}(t)+ t \left( t^2 - 1\right)\left[t u'_{k-1}(t) +
\frac{1}{2} u_{k-1}(t)\right]$

\hspace{175pt}($k =$ 1, 2, ...)

Using MAPLE, we have calculated these functions up to $v_{11}$,
although generating higher order expressions is trivial, as is the
case with $u_i$. We give these results below.

\begin{eqnarray}
v_1&=\frac {7}{24}{t}^{3}-\frac{3}{8}t\nonumber
\end{eqnarray}
\begin{eqnarray}
 v_2&=-\frac
{455}{1152}{t}^{6}+\frac {33}{64}{t}^{4}-\frac {15}{
128}{t}^{2}\nonumber
\end{eqnarray}
\begin{eqnarray}
v_3&=\frac {95095}{82944}{t}^{9}-\frac {6545}{3072}{t}^{7}+\frac {
5577}{5120}{t}^{5}-\frac {105}{1024}{t}^{3}\nonumber
\end{eqnarray}
\begin{eqnarray}
v_4&=-\frac {40415375}{7962624}{t}^{12}+\frac
{2739737}{221184}{t}^ {10}-\frac {2448017}{245760}{t}^{8}+\frac
{114439}{40960}{t}^{6}-\frac {4725}{32768}{t}^{4}\nonumber
\end{eqnarray}
\begin{eqnarray}
v_5&=\frac {5763232475}{191102976}{t}^{15}-\frac
{215656441}{2359296}{t}^{13}+\frac
{355886245}{3538944}{t}^{11}-\frac {280397117}{
5898240}{t}^{9}+\frac {15602073}{1835008}{t}^{7}-\frac {72765}
{262144}{t}^{5}\nonumber
\end{eqnarray}
\begin{eqnarray}
v_6&=-\frac {6183948445675}{27518828544}{t}^{18}+\frac
{415138648925}{ 509607936}{t}^{16}-\frac
{4775249765}{4194304}{t}^{14}+\frac {
7176153985}{9437184}{t}^{12}-\frac {75861726551}{314572800}{t}^
{10}+\nonumber
\end{eqnarray}
\begin{eqnarray}
&\frac {440748681}{14680064}{t}^{8}-\frac
{2837835}{4194304}{t}^{6} \nonumber
\end{eqnarray}
\begin{eqnarray}
v_7&=\frac {1329548915820125}{660451885056}{t}^{21}-\frac {
623575990562525}{73383542784}{t}^{19}+\frac {117495020467825}{
8153726976}{t}^{17}-\frac {11287669528993}{905969664}{t}^{15}+
\nonumber
\end{eqnarray}
\begin{eqnarray}
&\frac {4806755210517}{838860800}{t}^{13}-\frac {23169978705569}{
17616076800}{t}^{11}+\frac {28375388975}{234881024}{t}^{9}-\frac
{66891825}{33554432}{t}^{7}\nonumber
\end{eqnarray}
\begin{eqnarray}
v_8&=-\frac {2671063771882631125}{126806761930752}{t}^{24}+\frac {
59582343274078625}{587068342272}{t}^{22}-\frac {237670164192005545
}{1174136684544}{t}^{20}+\frac
{42077740772116621}{195689447424}{t}^{18}-\nonumber
\end{eqnarray}
\begin{eqnarray}
&\frac {1257093219318079}{9663676416}{t}^{16}+\frac {
296916207863309}{6710886400}{t}^{14}-\frac
{29041565208893}{3758096384}{t}^{12}+\frac{146540630595}{268435456}{t}^{10}-
\frac {14783093325}{2147483648}{t}^{8}\nonumber
\end{eqnarray}
\begin{eqnarray}
v_9&=\frac{6904699850316601458125}{27390260577042432}{t}^{27}-\frac
{ 461679745093525633675}{338151365148672}{t}^{25}+\frac {
29412227933816172445}{9393093476352}{t}^{23}-\nonumber
\end{eqnarray}
\begin{eqnarray}
&\frac { 37073088786771732695}{9393093476352}{t}^{21}+\frac {
6190356955971173843}{2087354105856}{t}^{19}-\frac {
519996976064854883}{386547056640}{t}^{17}+\frac {
3996930023590772587}{11274289152000}{t}^{15}-\nonumber
\end{eqnarray}
\begin{eqnarray}
&\frac { 1468251292588911}{30064771072}{t}^{13}+\frac
{517406211757245}{ 188978561024}{t}^{11}-\frac
{468131288625}{17179869184}{t}^{9}\nonumber
\end{eqnarray}
\begin{eqnarray}
 v_{10}&=-\frac
{4464578923214714502823625}{1314732507698036736}{t}^{30}+\frac
{1491741571661309972478475}{73040694872113152}{t}^{28}-\frac
{286485860646564818213975}{5410421842378752}{t}^{26}+\nonumber
\end{eqnarray}
\begin{eqnarray}
&\frac {17416724745836133903185}{225434243432448}{t}^{24}-\frac
{2318810066751674077595}{33397665693696}{t}^{22}+\frac
{16487466760916641832507}{417470821171200}{t}^{20}-\nonumber
\end{eqnarray}
\begin{eqnarray}
&\frac {45614196450845087013377}{3246995275776000}{t}^{18}+\frac {
12736018679229356269}{4294967296000}{t}^{16}-\frac
{49042918914307396335}{148159191842816}{t}^{14}+\frac
{22818897912239625}{1511828488192}{t}^{12}\nonumber
\end{eqnarray}
\begin{eqnarray}
&-\frac{33424574007825}{274877906944}{t}^{10}\nonumber
\end{eqnarray}
\begin{eqnarray}
v_{11}&=\frac
{1604407316678887857241980875}{31553580184752881664}{t}^{33}
-\frac {392956135061308232223935125}{1168651117953810432}{t}^{31}+
\frac
{1136426675054854043018733575}{1168651117953810432}{t}^{29}-\nonumber
\end{eqnarray}
\begin{eqnarray}
&\frac {209347382482809784715977475}{129850124217090048}{t}^{27}+
\frac {4057208263006803008962871}{2404631929946112}{t}^{25}-\frac
{7721779642093423553411219}{6679533138739200}{t}^{23}+\nonumber
\end{eqnarray}
\begin{eqnarray}
&\frac{
24317219180863821793468459}{46756731971174400}{t}^{21}-\frac{
185223434015774166216919}{1236950581248000}{t}^{19}+\frac{
109880525895631410224729}{4233119766937600}{t}^{17}-\nonumber
\end{eqnarray}
\begin{eqnarray}
&\frac{ 258444671539364377513}{107752139522048}{t}^{15}+\frac{
28529686078127770725}{314460325543936}{t}^{13}-\frac{
1327867167401775}{2199023255552}{t}^{11}\nonumber
\end{eqnarray}
\\
In Table \ref{table: derivnu300} below, we compare the values of
$J_{\nu}^{\prime}(x)$ obtained from the expansion with the exact
values, for $\nu = 300$. A similar table is not given for $\nu =
50000$ since we are not able to provide exact values for
comparison, due to the large value of the index.

\begin{table}[h]
\caption{Comparison of uniform expansion of $J_{\nu}^{\prime}(x)$
with exact values for $\nu = 300$} \vspace{10pt} \centering \small
\begin{tabular}{|c|c|c|c|c|c|c|} \hline \textbf{\textit{x}} &
\textbf{290} & \textbf{295} & \textbf{298} & \textbf{302} &
\textbf{305} & \textbf{310}\\ \hline \textbf{Uniform} & 0.0021909
& 0.0059599 & -0.010098 & -0.019458 & 0.0014413 & -0.018938\\
\hline \textbf{Exact} & 0.0021909 & 0.0059597 & 0.0084469 &
0.008023 & 0.0014411 & -0.018938\\ \hline
\end{tabular}\\
\label{table: derivnu300}
\end{table}

\section{Discussion}

We find that the uniform expansions are accurate to a higher
domain of validity than the expansions considered in our earlier
work \cite{CVRSW}, thereby increasing their utility especially in
the problematic $z \sim 1$ or ``transition" region. We compare the
uniform expansion with the exact Bessel function values in the
transition region in Figures \ref{fig2} and \ref{fig3} at the end
of this section, which are plotted for order $\nu=300$. It is seen
that the uniform expansion agrees with the exact Bessel function
curve for the entire range of the argument.

\begin{figure}[h]
\begin{minipage}[b]{0.47\textwidth}
\includegraphics[totalheight = 3 in, angle=270]{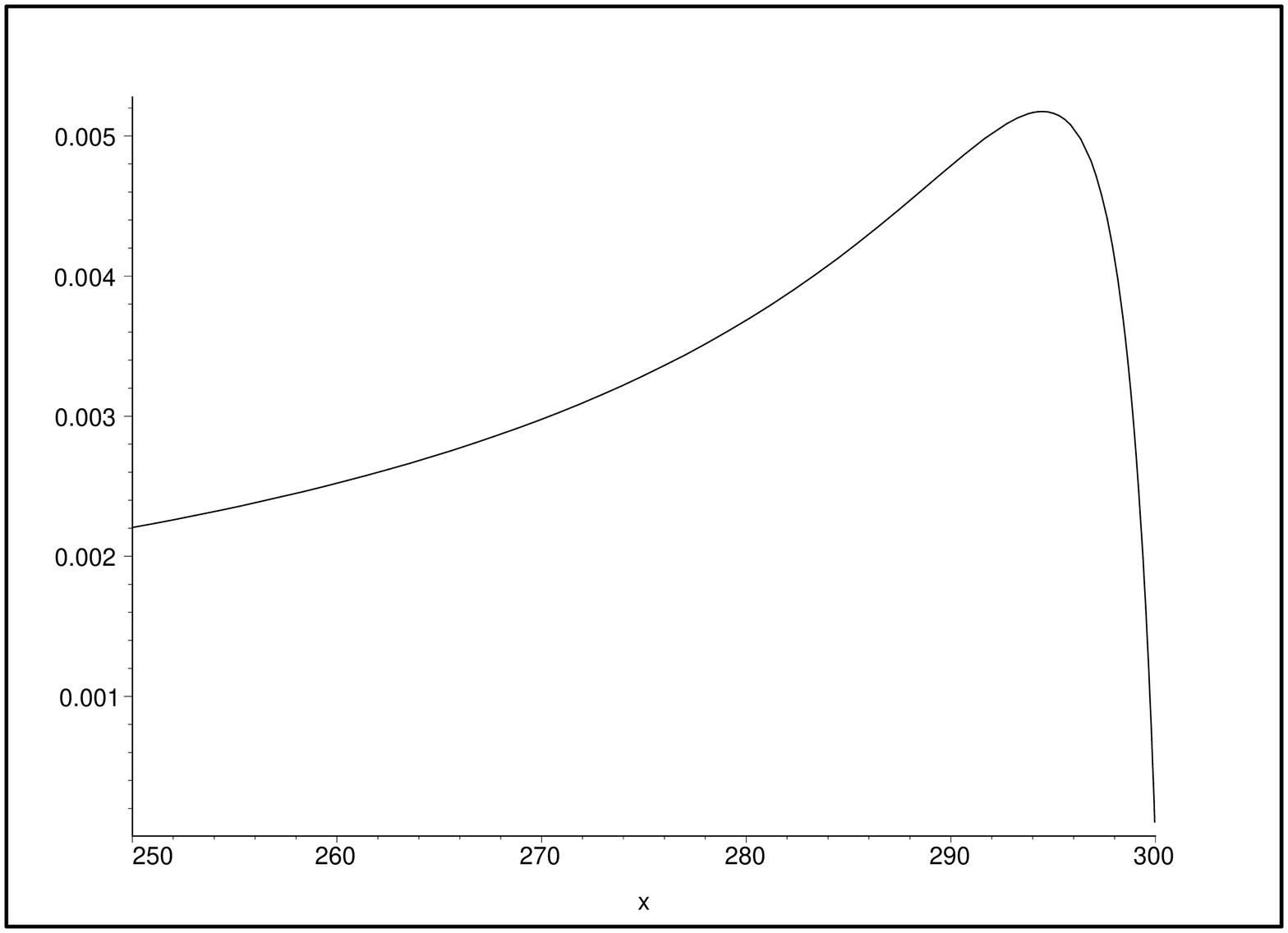}\\
\centering (a)
\end{minipage}
\begin{minipage}[b]{0.47\textwidth}
\includegraphics[totalheight = 3 in, angle=270]{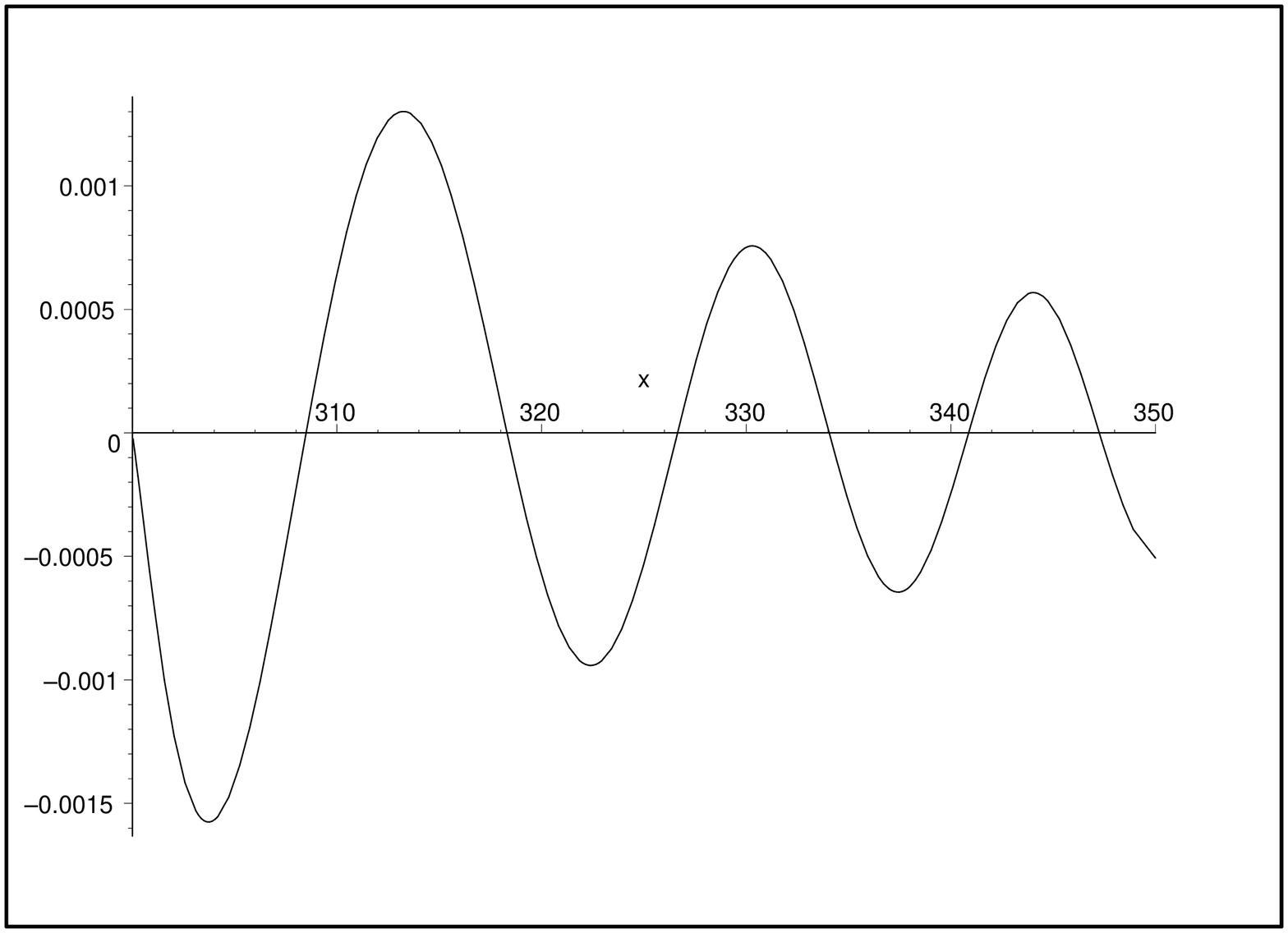}\\
\centering (b)
\end{minipage}
\caption{Plots of calculated values of (a) $\theta_1$ and (b)
$\theta_2$ vs. the argument of the Bessel function, for order $\nu
= 300$.}
\label{fig1}
\end{figure}

\begin{figure}[t]
\begin{minipage}{0.47\textwidth}
\centering \epsfig{file=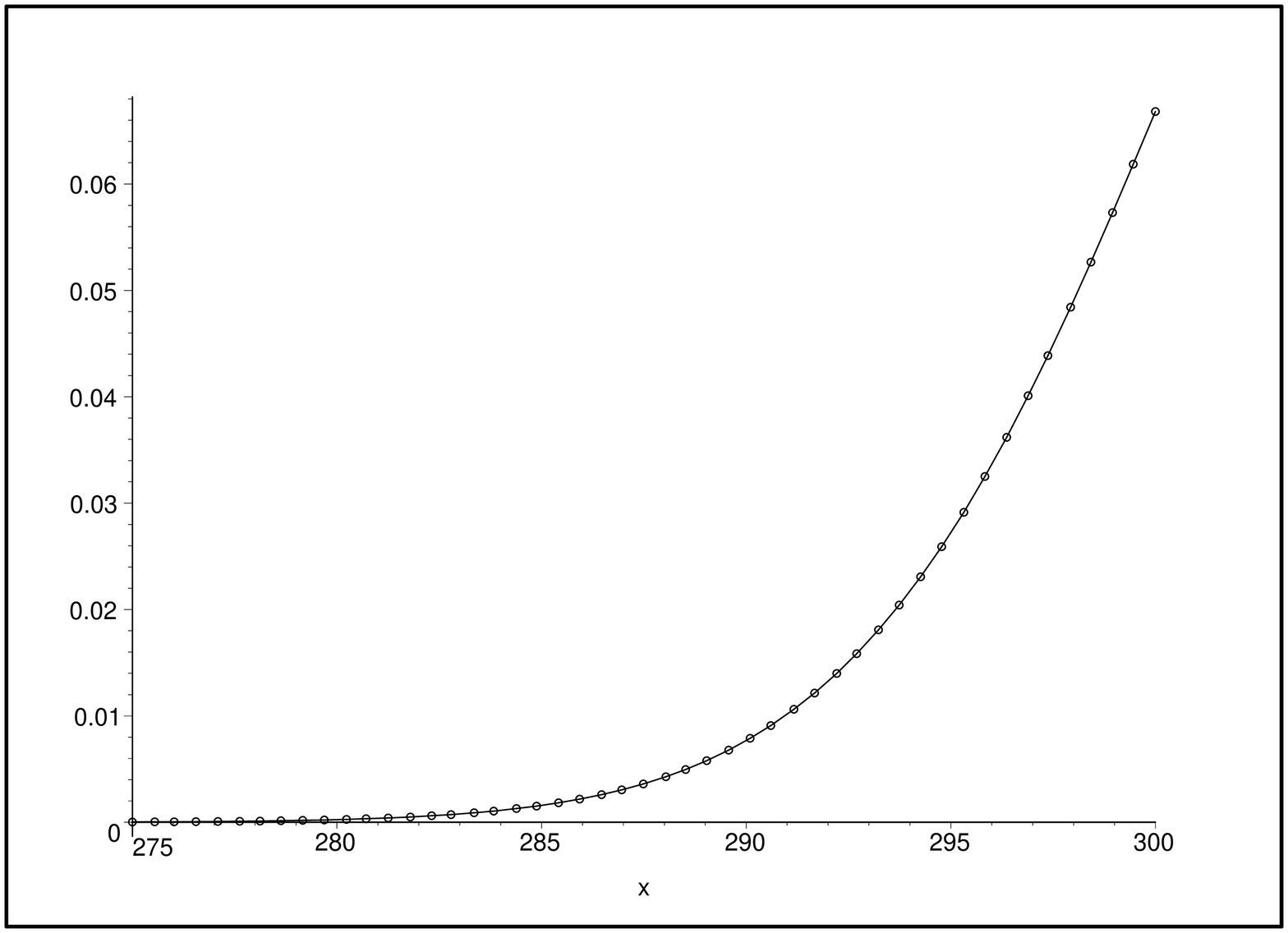,
width=0.7\textwidth, angle=270} \caption{Uniform Bessel expansion
and actual Bessel function graphed for argument $x$ and order
$\nu=300$ near the transition region. Solid line indicates actual
Bessel function values and circles indicate values given by the
expansion.} \vspace{50pt} \label{fig2}
\end{minipage}
\hspace{25pt}
\begin{minipage}{0.47\textwidth}
\centering \epsfig{file=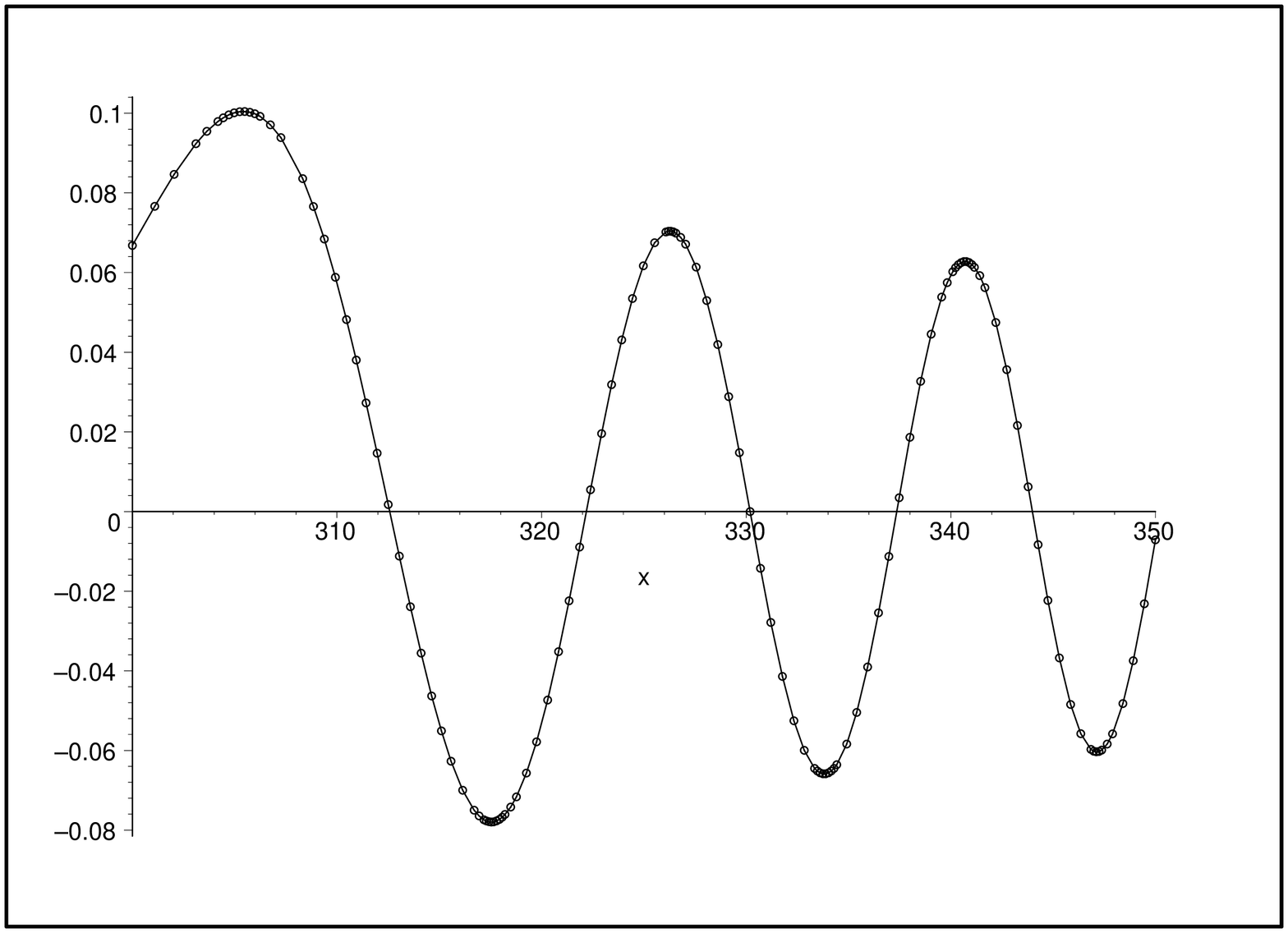, width=
0.7\textwidth, angle=270} \caption{Uniform Bessel expansion and
actual Bessel function graphed for argument $x$ and order
$\nu=300$ near the transition region. Solid line indicates actual
Bessel function values and circles indicate values given by the
expansion.} \vspace{50pt} \label{fig3}
\end{minipage}
\vspace{48pt}
\begin{minipage}{0.47\textwidth}
\centering \epsfig{file=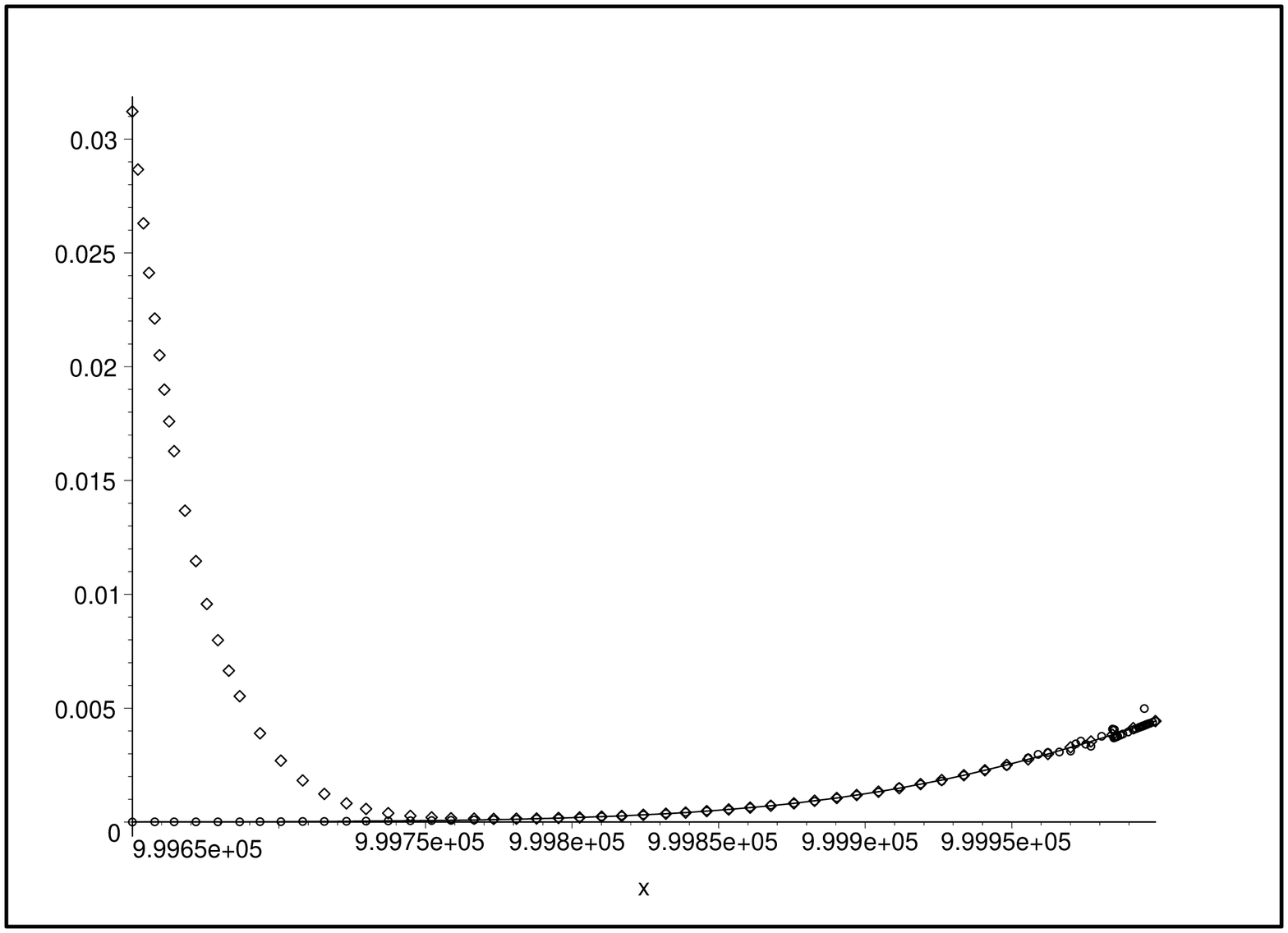,
width=0.7\textwidth, angle=270} \caption{Comparison of uniform
expansion, $\epsilon$ expansion and Watson's formulae for argument
$x<1,000,000$ and order $\nu=1,000,000$. Solid line indicates
uniform Bessel expansion values, diamonds represent $\epsilon$
expansion and circles indicate values given by Watson's formula.}
\label{fig4}
\end{minipage}
\hspace{25pt}
\begin{minipage}{0.47\textwidth}
\centering \epsfig{file=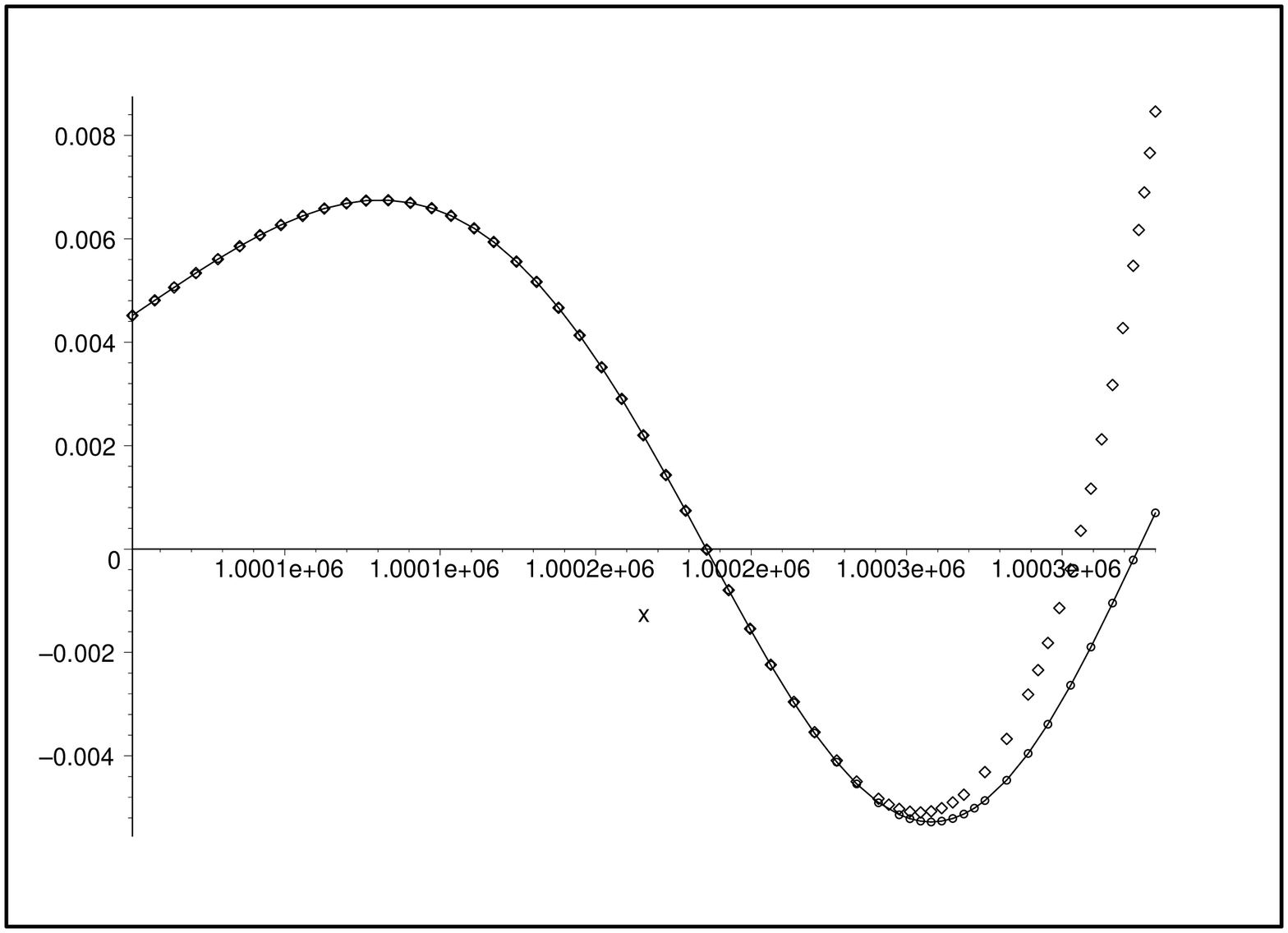,
width=0.7\textwidth, angle=270} \caption{Comparison of uniform
expansion, $\epsilon$ expansion and Watson's formulae for argument
$x>1,000,000$ and order $\nu=1,000,000$. Solid line indicates
uniform Bessel expansion values, diamonds represent $\epsilon$
expansion and circles indicate values given by Watson's formula.}
\label{fig5}
\end{minipage}
\end{figure}

Previously, we presented two methods that are geared to work
specifically in the transition region \cite{CVRSW}. First, we
present the results by Watson \cite{Watson}. For the case of the
argument being less than the order, he obtained, through the use
of contour integration,
\begin{eqnarray}
J_{\nu}(\nu sech(\alpha))= \frac{\tanh{\alpha}}{\pi
\sqrt{3}}\exp\left[\nu\left(\tanh{\alpha}+\frac{1}{3}\tanh^3{\alpha}-\alpha\right)\right]
K_{\frac{1}{3}}\left(\frac{1}{3}\nu\tanh^3{\alpha}\right)\nonumber
\end{eqnarray}
\begin{eqnarray}
\hspace{225pt}+3\theta_{1}\nu^{-1}\exp[\nu(\tanh{\alpha}-\alpha)]
\end{eqnarray}
where $|\theta_1|<1$. Similarly, for the case when the argument is
greater than the order, he derived the following:
\begin{eqnarray}
J_{\nu}(\nu\sec{\beta})=&
\hspace{-15pt}\frac{1}{3}\tan{\beta}\cos\left[\nu\left(\tan{\beta}-
\frac{1}{3}\tan^3{\beta}-
\beta\right)\right]\left(J_{-\frac{1}{3}}+J_{\frac{1}{3}}\right)&\nonumber\\
&+3^{-\frac{1}{2}}\tan{\beta}\sin\left[\nu\left(\tan{\beta}-\frac{1}{3}\tan^3{\beta}-\beta\right)\right]
\left(J_{-\frac{1}{3}}-J_{\frac{1}{3}}\right)&+24\theta_{2}\nu^{-1}
\end{eqnarray}

where $|\theta_2|<1$ and the argument for the Bessel functions
$J_{\pm\frac{1}{3}}$ is $\frac{1}{3}\nu\tan^{3}{\beta}$. The great
advantage of these formulae is that they have error bounds given.
We have found that $|\theta_1|$ and $|\theta_2|$ do not generally
have constant values for different values of the argument, but
they do nevertheless stay below 1. We have plotted the values of
$\theta_1$ and $\theta_2$ such that Watson's expansion matches the
exact value, for a range of arguments with $\nu = 300$ (Figure
\ref{fig1}). From these graphs, it is clear that $|\theta_1|$
stays below $\sim$ 0.0052 and $|\theta_2|$ stays below $\sim$
0.0016.

%\vspace{20pt}

%\vspace{20pt}

On the other hand, Debye \cite{DBY} introduced what we will term
as the ``$\epsilon$ expansion". The idea is motivated by
introducing a small parameter $\epsilon$, such that
$\nu=x(1-\epsilon)$, where $\nu$ denotes the order and $x$ is the
argument of the Bessel function. This expansion is given below:
\begin{equation}
J_{\nu}(x)\sim\frac{1}{3\pi}\sum_{m=0}^{\infty} B_{m}(\epsilon x)
\sin \left[ \frac{1}{3}(m+1)\pi \right] \cdot
\frac{\Gamma(\frac{1}{3}m+\frac{1}{3})}{(\frac{1}{6}x)^{(m+1)/3}}
\end{equation}

We extended this analysis of Debye by 5 orders to the terms
$B_m(\epsilon x), m=0,1,2,..15$ \cite{CVRSW}. It should be
remarked that the single value of the Bessel function when the
argument equals the order ($x=\nu$) was also considered in this
earlier study which extended the previous work of \cite{AIR} by
two orders. We observed that inclusion of the higher order terms
leads to highly accurate values, and so the singular case, for the
uniform expansions (when $x=\nu$) is well under control.

Now, to illustrate the applicability of these three methods to the
transition region ($z \sim 1$), we present Figures \ref{fig4} and
\ref{fig5}, which are plotted for the problematic regions relevant
for GW phenomenology (when the order is as large as
$\nu=1000000$). Watson's expansion and the $\epsilon$ expansion
both show remarkable ability in capturing the functions in the
transition region.

\begin{figure}[h]
\centering \epsfig{file=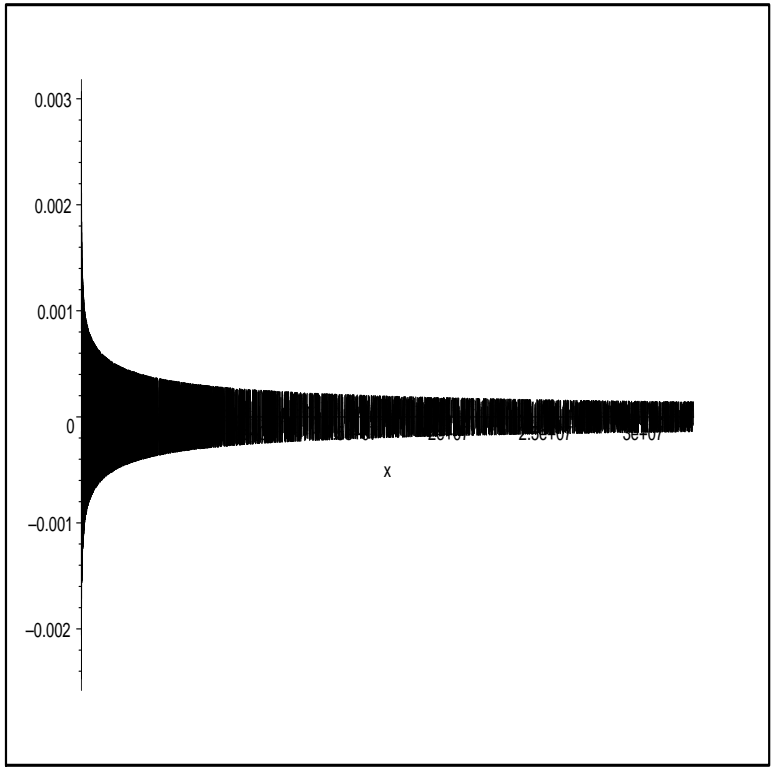,width=
4 in, height= 4 in, angle=270} \caption{Plot of uniform Bessel
expansion for argument $x$ ranging from 1,000,200 to 32,500,000
for order 1,000,000.} \label{fig6}
\end{figure}

We were not able to make exact comparison for such a large order,
obviously due to massive computer times required. However, in
Figures \ref{fig4} and \ref{fig5}, we observe strong evidence that
the proposed uniform Bessel asymptotic expansions are appropriate
for GW signal analysis. Here, we note the transition region
starting at values of the argument at 1,000,000 and going up to
1,000,200. In this region, the $\epsilon$ expansion, Watson's
formula and the uniform expansion closely coincide with one
another. As we found earlier \cite{CVRSW} the $\epsilon$ expansion
breaks down earlier as the argument leaves the transition region;
however, all three methods coincide in a certain region indicating
that we have consistent methods that work for values relevant to
GW analysis. The uniform Bessel expansion is fairly easy to
implement computationally and indicates good stability for rather
large values of the argument. To illustrate the type of values a
GW pulsar FT would require, we present Figure \ref{fig6}. Here, we
plot the uniform Bessel expansion for values ranging from
1,000,200 to 32,500,000, which are relevant for GW phenomenology.
This appears as a black band and is a continuous function which
indicates oscillations tightly bunched together, and happens to be
identical to a graph obtained earlier by using Meissel's second
expansion \cite{CVRSW}. It is noteworthy that the method is stable
and shows consistent behaviour over an extreme range of values for
the argument.

The zeroes (roots) of the Bessel function will help to identify
the sky locations in terms of $\theta$ where the signal strength
is zero. Zeroes of Bessel functions arise in a variety of
diffraction problems. Derivatives of Bessel functions are also
important in a multitude of applications and are evaluated using
the uniform expansions.

It is worth mentioning here a procedure developed by Baruth
\cite{Baruth} which can evaluate Bessel functions for large index
and argument, for example in the range of 5 million. The procedure
performs evaluations in less than 10 ms, and the author believes
that it can be improved even further. As an example, he has
calculated $J_{1000000}(999995) = 4.267834146855037\times
10^{-3}$, which is close to our value of $4.267834146855055\times
10^{-3}$, an agreement of 13 decimal places.
\\

\noindent {\bf Number-theoretic study of the denominators of
$u_i$}\\

We note here an interesting pattern regarding the functions $u_i$
used in the uniform expansion. Using MAPLE we undertook a study of
the denominators occurring in the $u_i$ mainly because their prime
number factorizations appear to have some surprising properties.

Let $d_i$ denote the common denominator of the terms of $u_i$ for
$i=0,1,\ldots$ and consider the prime number factorizations of the
first eleven $d_i$'s :

\begin{equation}
\begin{array}{cc}
i & d_i \\
\hline \\
0 & 1 \\
1 & 2^3 \, 3 \\
2 & 2^7 \, 3^2 \\
3 & 2^{10} \, 3^4 \, 5 \\
4 & 2^{15} \,3^5 \,5 \\
5 & 2^{18} \, 3^6 \, 5 \, 7 \\
6 & 2^{22} \, 3^8 \, 5^2 \, 7 \\
7 & 2^{25} \, 3^9 \, 5^2 \, 7 \\
8 & 2^{31} \, 3^{10} \, 5^2 \, 7 \\
9 & 2^{34} \, 3^{13} \, 5^3 \, 7 \, 11 \\
10 & 2^{38} \, 3^{14} \, 5^3 \, 7^2 \, 11 \\
\end{array}
\label{primeNumberFactorizations}
\end{equation}

\noindent {\bf Powers of $2$} \\
Let $a_i$ denote the exponent of $2$ in the prime number
factorizations of $d_i$ for $i=0,1,\ldots$. Then the following
formula for $a_i$ seems to be valid
\begin{equation}
    a_i = \log_2 \left( \frac{2^{4i}}{g_i} \right)
    \label{log2GouldFormula}
\end{equation}
where $g_i$ is Gould's sequence, see \cite{TCS92}, defined as the
highest power of $2$ dividing the central binomial coefficient
${2i \choose i}$. All terms of Gould's sequence $g_i$ are powers
of two and therefore $a_i$ will be an integer. This remarkable
identification was done via N. J. A. Sloane's {\em On-Line
Encyclopedia of Integer Sequences}, see \cite{NJAS}.

The table below lists the first eleven values given by formula
(\ref{log2GouldFormula}) which agree with the values given in the
previous table (\ref{primeNumberFactorizations}).
$$
\begin{array}{c||ccccccccccc}
i    & 0    & 1   & 2   & 3   & 4   & 5   & 6   & 7   & 8   & 9   & 10  \\
\hline
g_i  & 2^0  & 2^1 & 2^1 & 2^2 & 2^1 & 2^2 & 2^2 & 2^3 & 2^1 & 2^2 & 2^2 \\
\hline
a_i  & 0    &  3  & 7   & 10  & 15  & 18  & 22  & 25  & 31  & 34  & 38  \\
\end{array}
$$

\noindent {\bf Powers of $3$} \\
Let $b_i$ denote the exponent of $3$ in the prime number
factorizations of $d_i$ for $i=0,1,\ldots$. Then the following
formula for $b_i$ seems to be valid
\begin{equation}
    b_i = i + \mbox{ exponent of the highest power of } 3 \mbox{ dividing
} i!
    \label{powersOf3Formula}
\end{equation}

The table below lists the first eleven values given by formula
(\ref{powersOf3Formula}) which agree with the values given in
(\ref{primeNumberFactorizations}).
$$
\begin{array}{c||ccccccccccc}
i    & 0  & 1 & 2 & 3 & 4 & 5 & 6 & 7 & 8  & 9  & 10  \\
\hline
\mbox{exponent }  & 0  & 0 & 0 & 1 & 1 & 1 & 2 & 2 & 2  & 4  & 4 \\
\hline
b_i  & 0  & 1 & 2 & 4 & 5 & 6 & 8 & 9 & 10 & 13 & 14  \\
\end{array}
$$

These number theoretic properties could potentially be useful
toward finding an analytic solution to the integro-differential
equation describing the Bessel function for the uniform asymptotic
expansions.

\clearpage

\clearpage

\section{Conclusions}

In this work we examined, using symbolic computation, the accuracy
and the range of validity of uniform asymptotic expansions of
Bessel functions relevant for GW signal from pulsars and found
them to be valid over the entire domain of the argument, including
the problematic ``transition region". We compared these expansions
with the fractional transition Bessel and the Debye expansions in
such regions and found all three expansions to be applicable in
the transition region. It was shown that there is a region where
all three expansions agree, which is indicative of the accuracy of
the uniform expansion in this domain.

The functions used by Bessel in 1824 in connection with planetary
motion have found innumerable applications as earlier studies by
Bernoulli, Lagrange, Carlini, Laplace, Poisson, Euler and many
others \cite{Watson} have shown. Further problems will also bring
the ubiquitous Bessel functions to more diverse applications. Such
applications may require accurate calculations of the functions
for large order and argument, and asymptotic analysis will also be
relevant for other special functions.

\section{Acknowledgements}
\paragraph{}We are deeply grateful to SHARCNET (Shared Hierarchical Academic
Research Computing Network) and NSERC for grant support. We are
indebted to Dr. Nico Temme (CWI, Amsterdam) for suggesting the
importance of this study and for his invaluable advice on the
evaluation of the uniform expansion, and also to Dr. Dan Baruth
for bringing to our attention his work on evaluation of Bessel
functions of large order and argument.

\end{document}